\documentclass[twocolumn,showpacs,amsmath,amssymb,pra]{revtex4}

\usepackage{amsmath}
\usepackage{braket}
\usepackage{placeins}
\usepackage{bm}
\usepackage{color}
\usepackage{graphicx}

\newcommand{\ri}{{\rm i}}

\renewcommand{\H}{\hat{H}}
\newcommand{\Hdiag}{\hat{H}_{0}}
\newcommand{\Htun}{\hat{H}_{\text{tun}}}
\newcommand{\HBH}{\hat{H}_{\text{BH}}}

\newcommand{\creation}[1]{\hat{b}^\dag_{#1}}
\newcommand{\annihilation}[1]{\hat{b}^{\phantom{\dag}}_{#1}}
\newcommand{\n}[1]{\hat{n}^{\phantom{\dag}}_{#1}}
\newcommand{\JU}{{J}\big/{U}}
\newcommand{\muU}{{\mu}\big/{U}}

\begin{document}

\title[Hypergeometric analytic continuation]
      {Hypergeometric analytic continuation of the strong-coupling 
       perturbation series for the 2d~Bose-Hubbard model}

\author{S\"oren Sanders}
\email{soeren.sanders@uni-oldenburg.de}
	
\author{Christoph Heinisch}
\author{Martin Holthaus}

\affiliation{Institut f\"ur Physik, Carl von Ossietzky Universit\"at, 
        D-26111 Oldenburg, Germany}
	
\date{July 29, 2015}	

\begin{abstract}
We develop a scheme for analytic continuation of the strong-coupling  
perturbation series of the pure Bose-Hubbard model beyond the Mott 
insulator-to-superfluid transition at zero temperature, based on 
hypergeometric functions and their generalizations. We then apply this 
scheme for computing the critical exponent of the order parameter of this 
quantum phase transition for the two-dimensional case, which falls into the 
universality class of the three-dimensional $XY$~model. This leads to a
nontrivial test of the universality hypothesis. 
\end{abstract}



\maketitle
 

\section{Introduction}

In many branches of theoretical physics one encounters the necessity to
reconstruct an observable from a diverging perturbation series. This is
possible because a divergent series, by itself, is not a bad thing, but
still carries profound mathematical meaning~\cite{Euler60,Hardy49}. 
Take, for example, the geometric series
\begin{equation}
	\frac{1}{1-z} = \sum_{n=0}^\infty z^n \; ,
\end{equation}
where the sum on the right-hand side (r.h.s.) converges only for $|z| < 1$.
Nonetheless, if one regards the left-hand side (l.h.s.) as a definition of 
the formal sum, as already done by Euler~\cite{Euler60}, that sum obtains 
a well-defined meaning even for $|z| > 1$, giving, for instance,
$1 + 2 + 4 + 8 + \ldots = -1$.
%
%
Similarly, quantum-mechanical perturbation theory may yield a formal series
\begin{equation}
	A(\lambda) \sim \sum_{n=0}^\infty \alpha_n \lambda^n
\label{eq:PES}
\end{equation}	
for some quantity $A(\lambda)$, where $\lambda$ is a small parameter. 
The coefficients $\alpha_n$ may be such that the series is asymptotic, 
having zero radius of convergence, as it happens, e.g., when computing 
the ground-state energy of an anharmonic oscillator with a quartic 
perturbation of a quadratic potential~\cite{BenderWu69}. The task then 
again is to identify the unknown true observable on the l.h.s.\ from the 
given formal sum on the r.h.s.

The concept that naturally comes into play here is analytic continuation. 
Still, for applying this concept in practice, when merely a few leading
coefficients $\alpha_n$ are available, one needs to invoke some sort of
{\em a priori\/} hypothesis about $A(\lambda)$, either explicitly 
or implicitly. For instance, if one possesses explicit knowledge of 
$A(\lambda)$ for large values of $\lambda$, one can exploit this for 
designing rapidly converging strong-coupling expansions from divergent 
weak-coupling perturbation series~\cite{JankeKleinert95,KleinertEtAl97,
JaschKleinert01}. When resorting instead to the more familiar Pad\'e 
approximation technique~\cite{BakerMorris96,CalicetiEtAl07}, one introduces 
rational approximants of the form   
\begin{equation}
	A_{L/M}(\lambda) = 
	\frac{\sum_{n=0}^L p_n \lambda^n}{1 + \sum_{n=1}^M q_n \lambda^n}
\label{eq:RPA}
\end{equation}
and equates the coefficients obtained from a Taylor series expansion of
$A_{L/M}$ up to the order ${\mathcal O}(\lambda^{M+N})$ to those of the
perturbation series~(\ref{eq:PES}). While the resulting Pad\'e table then may 
yield good numerical values of the desired quantity, one is implictly imposing 
the asymptotic behavior $A_{L/M}(\lambda) \sim p_L \lambda^{L-M} / q_M$
for large $\lambda$, which may not be physically correct. 

Recently, Mera, Pedersen, and Nikoli\'c~\cite{MeraEtAl14} have suggested to 
replace the rational Pad\'e approximants~(\ref{eq:RPA}) by hypergeometric 
functions, which represent another form of an implicit {\em a priori\/} 
hypothesis. The examples from elementary single-particle quantum mechanics 
studied by these authors suggest that the corresponding analytic continuation 
technique can dramatically outperform Pad\'e and Borel-Pad\'e approaches. 
Hence, it was conjectured that the hypergeometric-function scheme might also 
be useful for many-body problems of condensed-matter physics\cite{MeraEtAl14}. 

In the present letter we provide first evidence which strongly supports this 
conjecture. We consider the two-dimensional (2d) Bose-Hubbard model on a 
square lattice, which constitutes a system of paradigmatic importance for 
quantum many-body theory~\cite{FisherEtAl89,FreericksMonien96,CapogrossoEtAl08,
SantosPelster09,FreericksEtAl09}, and develop a ``hypergeometric'' technique 
for obtaining the critical exponents of its Mott insulator-to-superfluid 
transition. 
      
We proceed as follows: We first briefly introduce the Bose-Hubbard model, 
and the strong-coupling perturbation series derived from it, which serves as 
input for the subsequent analysis. Next, we show how Gaussian hypergeometric 
functions $_2F_1$, and their generalizations $_{q+1}F_{q}$, emerge quite 
naturally when studying the quantum phase transition. We then apply our 
scheme for computing the phase diagram and the critical exponent of the 
order parameter. This is of some conceptual interest, since these 
exponents are supposed to be universal, but there is still a slight 
discrepancy~\cite{Vicari07} between experimentally measured~\cite{LipaEtAl03} 
and theoretically calculated~\cite{CampostriniEtAl01} values. Our scheme opens 
up a fresh approach to this subtle issue.

\section{The model} 

We consider the pure Bose-Hubbard model for interacting Bose 
particles~\cite{FisherEtAl89,FreericksMonien96,CapogrossoEtAl08,SantosPelster09,
FreericksEtAl09} on a two-dimensional square lattice at zero temperature. In 
its grand-canonical version this model is characterized by three parameters: 
The hopping matrix element~$J$, which quantifies the strength of the tunneling 
contact between neighboring lattice sites, the repulsion energy~$U$ provided 
by each pair of particles occupying the same lattice site, and the chemical 
potential~$\mu$. For a given value of $\mu$ the competition between particle 
delocalization due to tunneling and localization caused by repulsion leads 
to the well-known quantum phase transition from a Mott insulator to a 
superfluid when the ratio $J/U$ is gradually increased, starting from 
zero~\cite{FisherEtAl89}. We employ the Fock-space operators $\creation{i}$ 
and $\annihilation{i}$ which create or annihilate a Boson at the $i$th site, 
so that
\begin{equation}
 	\n{i} = \creation{i} \annihilation{i}
\end{equation}
counts the number of particles at that site, and use $U$ as the energy scale 
of reference. The non-dimensionalized Hamiltonian then is written as  
\begin{equation}
 	\HBH = \Hdiag + \Htun \; ,
\label{eq:HBH}
\end{equation}
where the site-diagonal part 
\begin{equation}
 	\Hdiag = \frac{1}{2} \sum_{i} \n{i}(\n{i}-1) - \muU \sum_{i} \n{i}
\label{eq:HDI}
\end{equation}
models the on-site repulsion and incorporates the chemical potential to fix
the particle number; this operator~(\ref{eq:HDI}) serves as the starting point 
for perturbative expansions~\cite{FreericksMonien96}. The further term 
\begin{equation}
 	\Htun = -\JU \sum_{\langle i,j\rangle} \creation{i} \annihilation{j} 
\end{equation}
accounts for the tunneling effect, with the sum ranging over all pairs of 
neighboring sites $i$ and $j$. Following Ref.~\cite{SantosPelster09}, 
we then break the particle-number conservation implied by this Bose-Hubbard 
model~(\ref{eq:HBH}) by adding spatially uniform sources and drains, 
\begin{equation}
 	\H = \HBH + \sum_i \eta (\creation{i} + \annihilation{i}) \; ,
\label{eq:EBH}
\end{equation}
where, without loss of generality, we have taken the dimensionless source 
strength~$\eta$ to be real. The quantity of interest now is the intensive 
ground-state energy
\begin{equation} 
	{\mathcal E}\!\left(\JU,\muU,\eta\right) = \langle \H \rangle / M \; ,
\end{equation}
where the expectation value is taken with respect to the ground state of
the extended model~(\ref{eq:EBH}), and $M$ denotes the number of sites, 
assumed to be so large that finite-size effects do not matter. From this we 
obtain the susceptibility 
\begin{equation}
	2 \psi = \frac{\partial {\mathcal E}}{\partial \eta}
	= 2 \langle \annihilation{i} \rangle \; ,
\end{equation}
where the first identity constitutes the definition of $\psi$, and 
the second is provided by the Hellmann-Feynman theorem. When taken at 
$\eta = 0$, this derivative describes the response of the original 
Bose-Hubbard model~(\ref{eq:HBH}) to the sources and drains: The expectation 
value $\langle \annihilation{i} \rangle = \psi$ is zero in the Mott-insulating 
phase, but takes on nonzero values in the superfluid phase, and thus serves
as order parameter. 

Assuming now that the ground-state energy per site can be expanded in
a power series of $\eta$, we write
\begin{multline}
	{\mathcal E}\!\left(\JU,\muU,\eta\right) \\ 
	= e_0\!\left(\JU,\muU\right) 
	+ \sum_{k=1}^\infty c_{2k}\!\left(\JU,\muU\right) \eta^{2k} \; .
\end{multline}
For each $\muU$ the coefficients $c_{2k}$, known as $k$-particle correlation 
functions, are then expanded in powers of $\JU$:
\begin{equation}
 	c_{2k}\!\left(\JU,\muU\right) = \sum_{\nu=0}^\infty 
	\alpha_{2k}^{(\nu)}\!\left(\muU\right) \left(\JU\right)^\nu \; .
\end{equation}
In order to make contact with the Landau theory of phase 
transitions~\cite{Landau37,Landau69,LaLifV}, the key idea now is to 
employ $\psi$ instead of $\eta$ as independent variable. This is achieved 
by means of a Legendre transformation, which leads to the effective
potential~\cite{SantosPelster09,BradlynEtAl09} 
\begin{eqnarray}
	\Gamma & = & {\mathcal E} - 2 \psi \eta
\nonumber \\	& = &
	 e_0 + a_2 \psi^2 + a_4 \psi^4 + a_6 \psi^6 + \mathcal{O}(\psi^8)
\label{eq:GAM}
\end{eqnarray} 
with the one-particle-irreducible vertices
\begin{equation}
 	a_2 = -\frac{1}{c_2} \; , \; 
	a_4 = \frac{c_4}{c_2^4} \; , \; 
	a_6 = \frac{c_6}{c_2^6}-\frac{4c_4^2}{c_2^7} \; , 
\end{equation}
having suppressed their dependence on $J/U$ and $\mu/U$. Since $\eta$ and
$\psi$ constitute a Legendre-conjugated pair, this construction implies
\begin{equation}
	\frac{\partial\Gamma}{\partial\psi} = -2\eta \; ,
\end{equation}
leading to the physical interpretation of the formalism: Since the actual
Bose-Hubbard system~(\ref{eq:HBH}) is recovered by setting $\eta = 0$, the
physical solutions correspond to the stable stationary points of 
$\Gamma$~\cite{SantosPelster09,BradlynEtAl09}.   

Now one can invoke a standard argument: Assuming $a_4$ and $a_6$ to be 
positive, and neglecting higher order terms of the expansion~(\ref{eq:GAM}), 
a single minimum of $\Gamma$ is found at $\psi_{\min} = 0$ as long as $a_2>0$, 
indicating the Mott insulator phase. In contrast, if $a_2<0$ the minimum is 
found at $\psi_{\min}\ne 0$, thus signaling the presence of the superfluid 
phase. Therefore, for given chemical potential $\mu/U$ the transition occurs 
when $a_2 = -1/c_2 = 0$, that is, at that value $(J/U)_{\rm c}$ at which the
series 
\begin{equation}
 	c_{2}\!\left(\JU,\muU\right) = 
	\sum_{\nu=0}^\infty \alpha_{2}^{(\nu)}\!\left(\muU\right) \,
	\left(\JU\right)^\nu
\label{eq:PSC}
\end{equation}
starts to diverge~\cite{TeichmannEtAl09a,TeichmannEtAl09b}. Moreover, from 
the usual Landau form $\Gamma \approx e_0 + a_2 \psi^2 + a_4 \psi^4$ one
obtains 
\begin{equation}
	\psi_{\rm min}^2 = \frac{-a_2}{2 a_4} 
\end{equation}	
for $J/U > (J/U)_{\rm c}$. Assuming $a_4$ to be positive and smooth at the 
transition, the exponent~$\beta$ which characterizes the emergence of the
order parameter according to 
\begin{equation}
	\psi_{\rm min}^2 \sim \left[ J/U - (J/U)_{\rm c} \right]^{2\beta}
\label{eq:DCE}
\end{equation}
is thus solely determined by $a_2 = -1/c_2$. This sets the stage for the 
present work: Its starting point is the perturbation series~(\ref{eq:PSC}) for 
the coefficient $c_2$. Although this series requires a small parameter $J/U$ 
it is referred to as a strong-coupling expansion~\cite{FreericksMonien96}, 
since it should converge in the strongly correlated Mott regime. We have
evaluated its coefficients $\alpha_2^{(\nu)}$ numerically up to the order
$\nu_{\rm max} = 10$ in $J/U$~\cite{TeichmannEtAl09a,TeichmannEtAl09b,
HinrichsEtAl13}, making use of the process-chain approach as devised in general
form by Eckardt~\cite{Eckardt09}. This technique, which has been recognized as 
an extremely powerful method~\cite{HeilVonderLinden12}, is based on Kato's 
non-recursive formulation of the Rayleigh-Schr\"odinger perturbation 
series~\cite{Messiah99}. Here we take these coefficients as input for 
determining optimal hypergeometric approximants to the Landau parameter 
$a_2$, as detailed in the following section, from which the respective 
exponents $\beta$ can then be read off directly.

\section{The Method} 

Given the coefficients $\alpha_2^{(\nu)}(\muU)$ for $\nu = 0$, $1$, $2$, 
\ldots, $\nu_{\rm max}$, the first task is to deduce the radius of convergence 
of the series~(\ref{eq:PSC}). This can be accomplished only if some 
{\em a priori\/} knowledge concerning the unknown higher-order coefficients 
is invested. To this end, useful guidance is provided by the case of high 
dimensionality~$d$: As explained in Ref.~\cite{TeichmannEtAl09b}, for 
$d \to \infty$ the expansion~(\ref{eq:PSC}) becomes a geometric series,
\begin{equation}
 	c_{2} = \alpha_2^{(0)} \sum_{\nu=0}^\infty
	\left(-2d\,\alpha_2^{(0)}\right)^\nu \left(\JU\right)^\nu \; ,
\end{equation}
from which one can immediately read off its radius of convergence
\begin{equation}
	(J/U)_{\rm c}= \frac{-1}{2d\,\alpha_2^{(0)}} \; ;
\end{equation}
after working out $\alpha_2^{(0)}$, this leads precisely to the mean-field
phase boundary~\cite{FisherEtAl89,TeichmannEtAl09b}
\begin{equation}
	(J/U)_{\rm c} = 
	\frac{(\muU + 1 - g)(g - \muU)}{2d\,(\muU + 1)} \; ,
\end{equation}
where the integer filling factor~$g$ satisfies $\muU + 1 \ge g \ge \muU$.
Consequently, in this limiting case the Landau coefficient $a_2 = -1/c_2$ 
takes the simple form
\begin{equation}
	a_2 = \frac{-1}{\alpha_2^{(0)}}
	\left( 1 - \frac{J/U}{(J/U)_{\rm c}} \right) \; ,
\end{equation}
exhibiting the mean-field exponent $2\beta = 1$.

\begin{figure}[tb]
\includegraphics[width=0.48\textwidth]{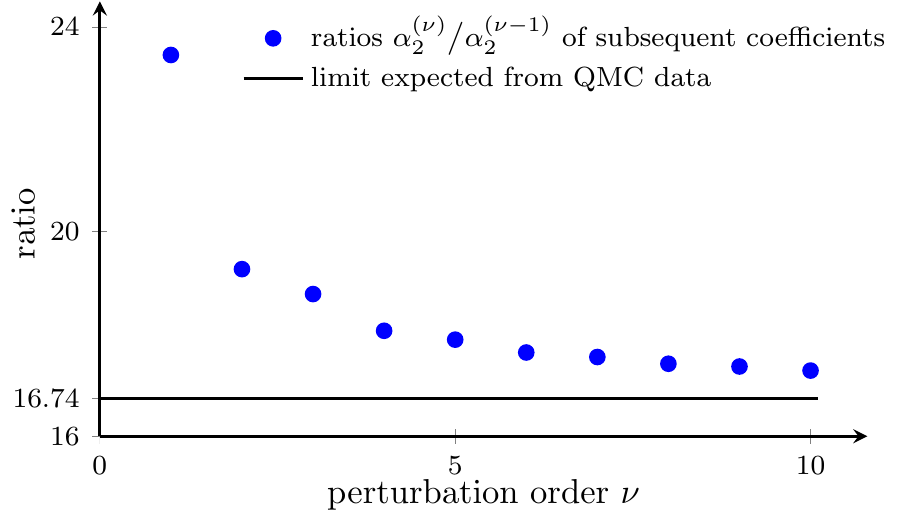}
\caption{Ratios ${\alpha_2^{(\nu)}}\big/{\alpha_2^{(\nu-1)}}$ of subsequent
	coefficients of the series~(\ref{eq:PSC}) for $d = 2$ and
	$\muU = 0.3769$, as corresponding to the tip of the lowest Mott lobe
	shown in Fig.~\ref{F_3}. The solid horizontal line indicates the
	limit $1/(J/U)_{\rm c}^{\rm QMC} = 1/0.05974(3)$ expected from QMC
	calculations~\cite{CapogrossoEtAl08}. Observe that all coefficients 
	$\alpha_2^{(\nu)}$ have the same sign, which impedes the Borel 
	summability of the series~(\ref{eq:PSC}).}
\label{F_1}
\end{figure}

For finite dimension~$d$, however, the ratio 
${\alpha_2^{(\nu)}}\big/{\alpha_2^{(\nu-1)}}$ of subsequent coefficients is 
not constant. For $d = 2$ this is shown exemplarily in Fig.~\ref{F_1} for 
$\mu/U = 0.3769$, corresponding to the tip of the lowest Mott lobe of the 
phase diagram depicted later in Fig.~\ref{F_3}. Thus, the corresponding 
series~(\ref{eq:PSC}) is no longer geometric, so that it is tempting to
{\em assume\/} a Landau coefficient of the form
\begin{equation}
	a_2 = \frac{-1}{\alpha_2^{(0)}}
	\left( 1 - \frac{J/U}{(J/U)_{\rm c}} \right)^{2\beta} \; , 
\end{equation}  
thereby admitting nontrivial exponents $2\beta$. According to this educated 
guess, the expansion~(\ref{eq:PSC}) should have the form of a binomial 
series,
\begin{equation}
	c_2 = \alpha_2^{(0)} \sum_{\nu=0}^\infty
	\frac{(2\beta)_\nu}{\nu!}
	\left( \frac{J/U}{(J/U)_{\rm c}} \right)^{\nu} \; ,
\label{eq:APA}
\end{equation}
where $(a)_\nu = a(a+1) \cdots (a+\nu-1)$ is the usual Pochhammer 
symbol~\cite{AbramowitzStegun72}. From an optimal fit of the given
coefficients~$\alpha_2^{(\nu)}$ to this hypothesis one could then 
determine approximate values of the two parameters $2\beta$ and
$(J/U)_{\rm c}$.

But this guess can still be improved: Realizing that the binomial series
coincides with the function $_1F_0$, employing the nomenclature used for
generalized hypergeometric functions~\cite{EMOT55}, one may generalize
the {\em a priori\/} ansatz~(\ref{eq:APA}) further and require
\begin{eqnarray}
	c_2 & = & \alpha_2^{(0)} 
	{_2F_1}\!\left(a,b;c;\frac{J/U}{(J/U)_{\rm c}}\right) 
\nonumber \\	& = &	
	\alpha_2^{(0)} 
	\sum_{\nu=0}^\infty \frac{(a)_\nu\,(b)_\nu}{\nu!\,(c)_\nu} 
	\left(\frac{J/U}{(J/U)_{\rm c}}\right)^\nu,
\label{eq:APB}
\end{eqnarray}
where now $_2F_1(a,b;c;z)$ denotes the well-known Gaussian hypergeometric
function~\cite{AbramowitzStegun72,EMOT55}, giving us four degrees of freedom 
for a least-square fit to the $\nu_{\rm max}$ perturbative data. The strength 
of its singularity at the point of divergence is given by
\begin{equation}
	{_2F_1}(a,b;c;z) \propto \frac{1}{(1-z)^{a+b-c}}
	\qquad \text{for } z \to 1 \; ,
\label{eq:SIN}
\end{equation}  
from which one finds the exponents
\begin{equation}
	2\beta = a + b - c \; .
\end{equation}
Going still one step further, one may replace $_2F_1$ by a generalized 
hypergeometric function $_{q+1}F_q$ providing $2q + 2$ degrees of freedom, 
requiring the evaluation of the perturbation series at least to the 
corresponding order. This possibility to perform analytic continuation of a 
perturbation series by means of an analytic function which itself is a member 
of a general familiy of ``higher order'' functions is the core of the
proposal made in Ref.~\cite{MeraEtAl14}. The hypergeometric functions, 
just as the binomial series, can adopt complex values beyond their radius of 
convergence. In view of its physical meaning we require $c_2$, as well as 
$a_2$, to be a real quantity, so that we take the real part of ${_{q+1}F_q}$. 
Technically, this is achieved by computing $\lim_{\varepsilon \to 0} 
\big({_{q+1}F_q}(x+\ri\varepsilon) + {_{q+1}F_q}(x-\ri\varepsilon)\big)/2$. 

Thus, from the expectation of a nontrivial exponent we infer that $c_2$ has 
to have an essential singularity at $(J/U)_{\rm c}$ with a strength determining 
the respective exponent. This is why $_{q+1}F_q$ are suitable approximants: As 
exemplified by Eq.~(\ref{eq:SIN}), the strength of their singularities can be 
tuned by adjusting their parameters.

\section{Results} 

\begin{figure}[tb]
\includegraphics[width=0.48\textwidth]{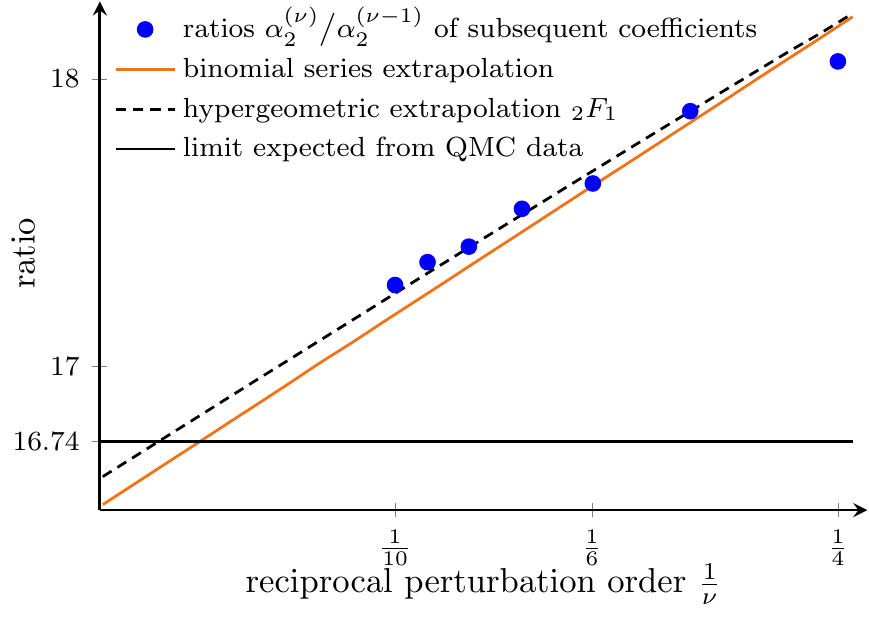}
\caption{Ratios ${\alpha_2^{(\nu)}}\big/{\alpha_2^{(\nu-1)}}$ of subsequent
	coefficients of the series~(\ref{eq:PSC}) for $d = 2$ and 
	$\mu/U = 0.3769$, plotted vs.\ the reciprocal order $1/\nu$, 
	together with the corresponding ratios obtained from the 
	fit~(\ref{eq:FBS}) to the binomial series, and from the 
	fit~(\ref{eq:FGH}) to the Gaussian hypergeometric function. Again, 
	the horizontal line indicates the limit 
	$1/(J/U)_{\rm c}^{\rm QMC} = 1/0.05974(3)$ expected from QMC
	calculations~\cite{CapogrossoEtAl08}.}
\label{F_2}
\end{figure}

We have applied this strategy to the series~(\ref{eq:PSC}) for the 
two-dimensional Bose-Hubbard model with $0 \le \mu/U \le 4$, having 
at disposal its coefficients $\alpha_2^{(\nu)}$ up to 
$\nu_{\max} = 10$~\cite{TeichmannEtAl09a,TeichmannEtAl09b,HinrichsEtAl13}. 
Figure~\ref{F_2}  again displays the ratios 
${\alpha_2^{(\nu)}}\big/{\alpha_2^{(\nu-1)}}$ of subsequent coefficients 
for $\mu / U =0.3769$, now plotted vs.\ the reciprocal order $1/\nu$. 
In addition, we also indicate by continuous lines the ratios resulting 
from the best fit  
\begin{equation}
	\alpha_2^{(\nu)} = \alpha_2^{(0)} \cdot 
	\frac{\big(1.405\big)_\nu}{\nu!} \cdot 16.51^{\nu}
\label{eq:FBS}	
\end{equation}
to the binomial series hypothesis~(\ref{eq:APA}), and from the best fit
\begin{equation}	
	\alpha_2^{(\nu)} = \alpha_2^{(0)} \cdot 
	\frac{\big(1.398\big)_\nu \, \big(-0.7684\big)_\nu}
	     {\nu! \, \big(-0.7606\big)_\nu} \cdot 16.61^\nu
\label{eq:FGH} 
\end{equation}
to the Gaussian hypergeometric hypothesis~(\ref{eq:APB}). Evidently, the 
quality of these fits is excellent. This allows us to perform reliable 
extrapolations to $\nu \to \infty$, where the ratios should approach the 
expected value $1/(J/U)_{\rm c}^{\rm QMC} = 1/0.05974(3)$ known from quantum  
Monte Carlo (QMC) simulations~\cite{CapogrossoEtAl08}. In addition, we have 
also fitted the exact coefficients to those of generalized hypergeometric 
functions $_3F_2$ and $_4F_3$.

\begin{figure}[tb]
\includegraphics[width=0.48\textwidth]{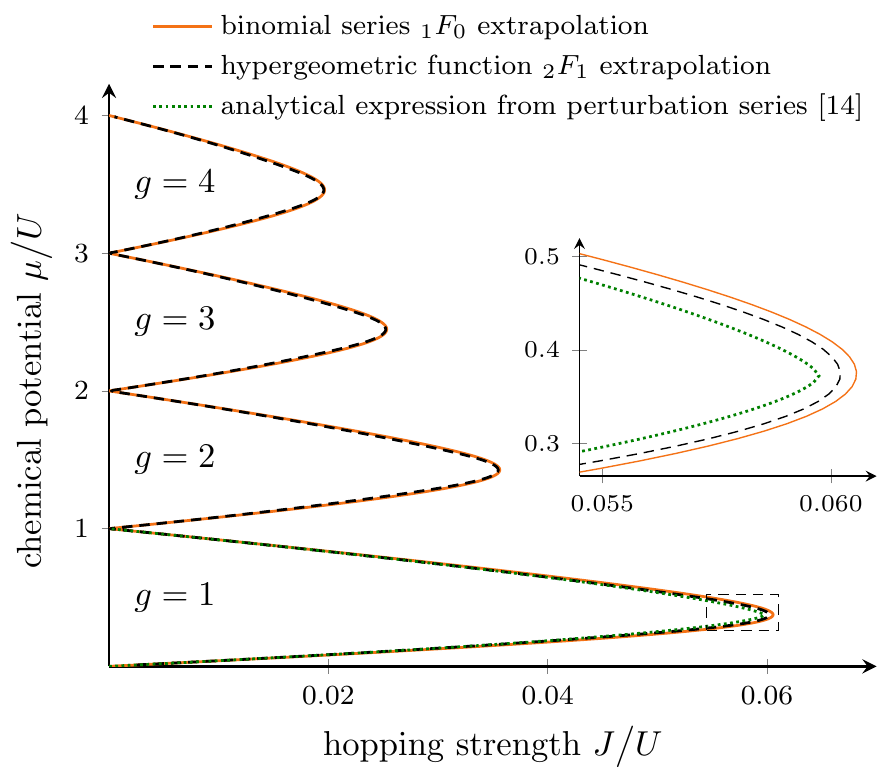}
\caption{Zero-temperature phase diagram of the 2d Bose-Hubbard model as 
	obtained from the binomial series hypothesis, and from the Gaussian 
	hypergeometric hypothesis. Inside the lobes the system is in a Mott 
	insulator state with $g$~particles per site, outside these lobes 
	in a superfluid state. Also shown is the expression for the lowest 
	Mott lobe stated in Ref.~\cite{FreericksEtAl09}.}
\label{F_3}
\end{figure}

Performing these procedures for all values of the chemical potential that
are of interest, and reading off the respective values of $(J/U)_{\rm c}$, 
we obtain the system's phase diagram. In Fig.~\ref{F_3} we show this
diagram for $0 \leq \mu/U \leq 4$, as resulting from the binomial and
from the Gaussian hypergeometric fit, respectively. The relative deviation 
between both curves stays below 4\%. It is largest halfway between the
position of a tip of a Mott lobe and the nearest integer values of $\mu/U$;
at the tips these deviations are smaller than~$1\%$.

The tips of the Mott lobes represent multicritical points with particle-hole
symmetry; here the system falls into the universality class of the 
$(d+1)$-dimensional $XY$~model~\cite{FisherEtAl89}. The critical scaled 
hopping strength~$(J/U)_{\rm c}$ at the tip of the lowest lobe for $d = 2$, 
as provided by the respective fit, figures as 
\begin{equation}
 \begin{aligned}
	{_1F_0}& \; : \; (J/U)_{\rm c} \; = \; 0.06056\\
	{_2F_1}& \; : \; (J/U)_{\rm c} \; = \; 0.06021\\
	{_3F_2}& \; : \; (J/U)_{\rm c} \; = \; 0.06003\\
	{_4F_3}& \; : \; (J/U)_{\rm c} \; = \; 0.06004\; ,
 \end{aligned}
\end{equation}
which matches the QMC value $(J/U)_{\rm c}^{\rm QMC} = 0.05974(3)$ quite
well. If we assume this value to be exact, and take the result provided
by $_2F_1$ as sound compromise between the number of coefficients available
and the number of fit parameters, the error of that result is less than 
$1\%$.

\begin{figure}[tb]
\includegraphics[width=0.48\textwidth]{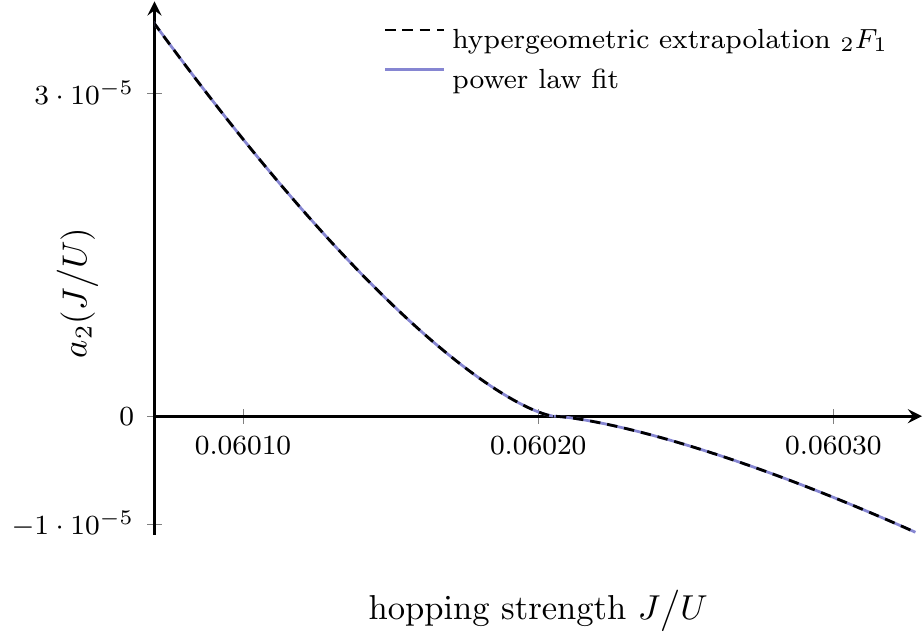}
\caption{Behavior of the Landau coefficient $a_2$ for $\mu/U = 0.3769$,
	as obtained by Gaussian hypergeometric continuation; for
	$J/U > (J/U)_{\rm c}$ the real part is shown. Also shown is
	the power-law fit~(\ref{eq:FIT}).}  
\label{F_4}
\end{figure}

So far, we have used hypergeometric continuation merely to reproduce existing 
knowledge, thus confirming its reliability. However, with the determinantion 
of the order parameter's exponent we now move to a ground which is technically
far more demanding~\cite{RanconDupuis11}. In Fig.~\ref{F_4} we show the 
Gaussian hypergeometric description of the Landau coefficient $a_2$, and its
analytic continuation beyond the transition point, again at the tip of
the lowest Mott lobe. The data are well described by the power-law fit 
\begin{equation}\label{eq:FIT}
 \begin{aligned}
	8.679\big( (J/U)_{\rm c} - J/U \big)^{1.390} 
	& \text{for} & J/U < (J/U)_{\rm c}\\
	-2.945\big( J/U - (J/U)_{\rm c} \big)^{1.390} 
	& \text{for} & J/U > (J/U)_{\rm c}.
 \end{aligned}
\end{equation}

\begin{figure}[tb]
\includegraphics[width=0.48\textwidth]{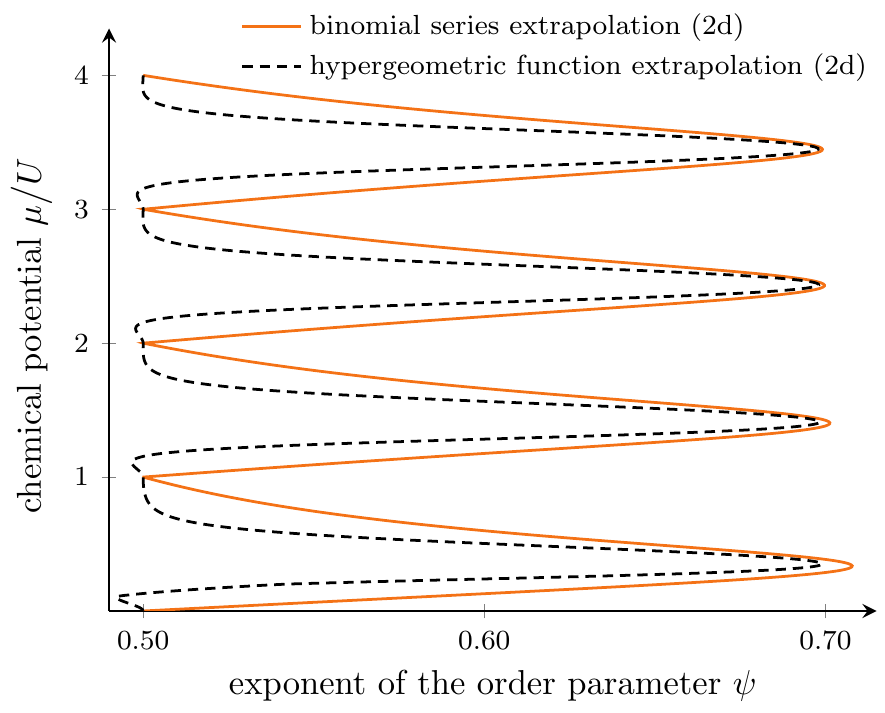}
\caption{Exponents $\beta$ of the order parameter $\psi$ for the
	Mott insulator-to-superfluid transition, as obtained by binomial
	and Gaussian hypergeometric continuation, respectively. The
	``critical'' value $\beta_c \approx 0.69$ is adopted at the
	tips of the Mott lobes.}
\label{F_5}
\end{figure}

Performing this procedure within the interval $0 \le \mu/U \le 4.0$ for both 
the binomial and the Gaussian hypergeometric ansatz, we obtain the exponents 
displayed in Fig.~\ref{F_5}. While both variants lead to notably different 
results for general $\mu/U$, they agree quite well at the tip of the lobes, 
{\em i.e.\/}, at the critical points. This observation is quite significant, 
since one expects nontrivial critical exponents only at the tips of the 
lobes, while the system should be mean-field like for all other $\muU$. 
Our approximate scheme can only yield continuous lines, and it is an open 
question whether these would reduce to $\delta$-like spikes with higher orders.
However, the relative stability of the data at the lobes' tips indicates that 
one can actually determine the true critical exponent $\beta_{\rm c}$ of the 
Mott insulator-to-superfluid transition with good accuracy by hypergeometric 
continuation. In particular, at the tip of the lowest lobe we obtain the 
following values:      
\begin{equation}
 \begin{aligned}
	{_1F_0}& \; : \; \beta_{\rm c} \; = \; 0.7023\\
	{_2F_1}& \; : \; \beta_{\rm c} \; = \; 0.6949\\
	{_3F_2}& \; : \; \beta_{\rm c} \; = \; 0.6881\\
	{_4F_3}& \; : \; \beta_{\rm c} \; = \; 0.6904 \; .
 \end{aligned}
\end{equation}
This gives rise to a nontrivial test of the universality hypothesis for 
critical phenomena. Using the scaling relation $\beta_c = (1 + \eta) \nu$, and 
inserting the best known estimates for the critical exponents $\eta = 0.0380(4)$
and $\nu = 0.67155(27)$, as derived from the three-dimensional $XY$ model by 
combining Monte Carlo simulations based on finite-size scaling methods and 
high-temperature expansions~\cite{CampostriniEtAl01}, the assumption of 
universality yields the expectation $\beta_{\rm c} = 0.6971(6)$. Indeed,
this coincides within less than 1\% with our $_2F_1$-estimate extracted from 
the 2d Bose-Hubbard model. It remains to be seen whether further refinement 
of our approach will result in an even better confirmation of the universality 
hypothesis.

\section{Conclusions} 

In summary, we haven taken up an idea put forward by Mera, Pedersen, and
Nikoli\'c, who have suggested to utilize hypergeometric functions for the 
analytic continuation of divergent perturbation series~\cite{MeraEtAl14};
here we have adapted this concept to the strong-coupling perturbation 
series~(\ref{eq:PSC}) of the Bose-Hubbard model. After evaluating this series 
to the maximum accessible order in the scaled hopping strength~$J/U$, which is 
$\nu_{\rm max} = 10$ in the present case, we are in a position to determine 
the parameters of its hypergeometic approximants from a least-square fit with 
high accuracy. This has enabled us to assess the critical exponent of the 
order parameter of the Mott insulator-to-superfluid transition. Compared to a  
previous attempt to deduce critical exponents from diverging perturbation 
series~\cite{HinrichsEtAl13}, the present approach is conceptually simpler, 
and more easy to handle in practice. The success of this approach indicates 
that hypergeometric functions, and their generalizations, indeed embody the 
proper {\em a priori\/} knowledge required by this quantum phase transition. 
Aside from further refinements, the next steps to be taken with hypergeometric 
analytic continuation will involve the investigation  of the superfluid 
density, and the corresponding analysis of the 3d~Bose-Hubbard model.


\begin{acknowledgments}
One of us (M.H.) wishes to thank D.~Hinrichs and A.~Pelster for long
discussions.
The computations were performed on the HPC cluster HERO, located at 
the University of Oldenburg and funded by the DFG through its Major 
Research Instrumentation Programme (INST 184/108-1 FUGG), and by the 
Ministry of Science and Culture (MWK) of the Lower Saxony State.		
\end{acknowledgments}


\end{document}